# Formation of autophagosomes coincides with relaxation of membrane curvature.

Running title: Theoretical modelling of autophagosome formation.

Jaime Agudo-Canalejo[1,2] and Roland L. Knorr[3,4]

[1] Rudolf Peierls Centre for Theoretical Physics, University of Oxford, Oxford OX1 3NP, United Kingdom

[2] Department of Chemistry, The Pennsylvania State University, University Park, Pennsylvania 16802, United States

[3] Max Planck Institute of Colloids and Interfaces, Department of Theory & Bio-Systems, 14424 Potsdam, Germany

[4] Graduate School and Faculty of Medicine, The University of Tokyo, Tokyo 113-0033, Japan

jaime.agudocanalejo@physics.ox.ac.uk, roland.knorr@mpikg.mpg.de

**Abstract**

Autophagy is an intracellular degradation process that employs complex membrane dynamics to isolate and break down cellular components. However, many unanswered questions remain concerning remodeling of autophagic membranes. Here, we focus on the advantages of theoretical modelling to study the formation of autophagosomes and to understand the origin of autophagosomal membranes. Starting from the well-defined geometry of final autophagosomes we ask the question of how these organelles can be formed by combining various preautophagosomal membranes such as vesicles, membrane tubules or sheets. We analyse the geometric constraints of autophagosome formation by taking the area of the precursor membranes and their internal volume into account. Our results suggest that vesicle fusion contributes little to the formation of autophagosomes. In the second part we quantify the curvature of the precursors and report that the formation of autophagosomes is associated with a strong relaxation of membrane curvature energy. This effect we find for a wide range of membrane asymmetries. It is especially strong for small distances between both autophagosomal membranes, as observed *in vivo*. We quantify the membrane bending energies of all precursors by considering membrane asymmetries. We propose that the generation and supply of pre-autophagosomal membranes is one limiting step for autophagosome formation.

**Key Words:** reverse autophagy, membrane remodeling, fusion, scission, bending energy, membrane curvature, theory modelling, retrograde autophagy

## 1 Introduction

During the intracellular process of macroautophagy (hereafter autophagy), a membrane bound organelle, the autophagosome, is generated *de novo*. The remodeling of the autophagic membrane during the life cycle of the organelle is a complex multistep process involving changes in the morphology and topology of the autophagic membranes[1]. Autophagosomes grow as flattened double-membrane sheets, the phagophores, which then bend into spherical double-membrane organelles, the autophagosomes. Canonical mechanisms of membrane bending such as protein coats[2] were assumed to mediate the bending step[3]. However, using theoretical modelling, we previously showed that bending of phagophores has to be understood in terms of two coupled morphological changes. Bending of the flat sections of the sheet competes with the removal of the strongly curved edge of the phagophore. We showed that the energy profile of the bending process strongly depends on the relative size of the phagophore. The edge removal and bending of the phagophore are energetically favored for typical sizes of phagophores/autophagosomes *in vivo*[4]. This suggests that phagophores require mechanisms to prevent premature bending rather than a machinery to drive their bending.

Nevertheless, the origin and remodelling of early autophagic membranes is still a matter of debate[5]. Atg9 and COPII vesicles as well as membrane tubules seem to play central roles during autophagy initiation[6]. In mammals, autophagosomes can start, for example at contact sites of organelles[7] or as PI(3)P-positive domains of the endoplasmic reticulum (ER) known as the omegasome[8]. In itself, the ER is a combination of tubular and sheet-like membranes[9]. In general, membrane vesicles, tubules or sheets and combinations thereof (including more complex shapes) could contribute membrane material to form autophagosomes, see Fig. 1. In contrast to complicated, pre-autophagosomal mechanisms and potential membrane geometries, the final autophagosomes are rather simple and well-defined in terms of geometry: two spherical vesicles stuck within each other. Starting from the final shape of an autophagosome, we calculate the size and number of preautophagosomal vesicles, and the size of membrane tubules and sheets required to match the geometric constraints of the autophagosome. We consider the combinations of precursors such as vesicles, tubules and sheets. Moreover, we study the energetics of autophagosome formation.

This chapter is structured as following: in Section 2, we describe the parameters required to calculate the bending energy and show where the model can be simplified. In Section 3 we describe the geometry and energetics of common membrane shapes and consider the formation of autophagosomes under area and volume constraints present. In Section 4 we compute the energetics of autophagosome formation.

## 2 Materials and Methods

All biological membranes are in a fluid state. Because of their small thickness, of the order of 5 nm, they can be considered as quasi-two-dimensional fluid materials. The physical properties of the membrane differ from other quasi-two-dimensional fluids we might be familiar with in our macroscopic world. Soap-bubbles for example, conserve their internal volume but not their surface area. However, both the enclosed volume and the area of biological membranes is conserved. Thus, it is difficult to predict the behaviour of membranes using intuition based on our macroscopic experience with two- or three-dimensional fluids. Here we describe modelling of membranes as a method to understand autophagic membrane dynamics. The shape of fluid membranes is primarily governed by their bending energy[10]. This mesoscopic description has been corroborated by a detailed and quantitative comparison between experimentally observed and theoretically calculated shapes[11,12].

### 2.1 The bending energy of the membrane

The bending energy of a membrane is given by

$$E_{\text{be}} = 2\kappa \int dA \, (M - m)^2 + \kappa_G \int dA \, K \quad (1)$$

1. The integration runs over the whole **surface area** $A$ of the membrane or vesicle.
2. The **curvature** of any surface can be described locally by two perpendicular arcs. The inverse radii of these arcs are the two principal curvatures, $C_1$ and $C_2$, which characterize the local shape of a membrane. The principal curvatures can be positive or negative if the membrane bulges towards the exterior or interior compartment of a vesicle, respectively. The arithmetic mean of both curvatures defines the mean curvature $M = (C_1 + C_2)/2$, whereas their product defines the Gaussian curvature $K = C_1 C_2$.

3. The bending energy of any material is governed by its **bending rigidity** (or elastic modulus) $\kappa$, and by its **Gaussian curvature modulus** $\kappa_G$, both of which have units of energy. Fluid bilayers are very flexible and their bending rigidity is on the order of $\kappa = 10 - 20\, k_B T \sim 10^{-19}$ J, where $k_B$ is the Boltzmann constant and $T$ the temperature. The Gaussian curvature modulus is typically negative with $\kappa_G \approx -\kappa$, and we will use this value throughout this work unless otherwise noted.
4. If the two leaflets of the bilayer membrane differ in their molecular composition, the membrane can develop a certain preferred or **spontaneous curvature** $m$. The composition of a membrane can change, e.g. by binding of small molecules or proteins, or by their modification such as through phosphorylation.
5. Due to the Gauss-Bonnet theorem, the second integral in (1) does not depend on the specific shape of the membrane, it only depends on the membrane topology. For any closed surface with spherical topology, such as a sphere, a tubule closed at both ends, a pancake-like membrane sheet, or an autophagosome (prior to the scission of the neck), this term evaluates simply to $4\pi\kappa_G$.

## 2.2 Simplifications of the model

Depending on the actual aim of the model, a number of simplifications can be included. Here we introduce some generalizations we apply later in our reductionist model, mostly related to the interaction of proteins with the pre-autophagosomal membranes.

7. Binding and unbinding of any molecules or structural changes of these molecules alter the spontaneous curvature $m$. The formation of the autophagosome is achieved by the concerted actions of Atg proteins. Eighteen of the core proteins are grouped into six distinct complexes and most of these core complexes and their interactions are conserved, and their interaction with the membrane can be restricted in space and time[13]. For example, *in vitro* we showed that conjugating the ubiquitin-like protein Atg8 to phospholipids changes $m$ [14]. However, it is unknown how $m$ changes dynamically during the process of autophagy. It will be challenging to quantify such values and moreover, these values may differ between various types of autophagy. Thus, throughout most of this chapter we do not take $m$ into account. When we do, we will consider a homogeneous value of $m$ throughout the membrane, that could arise for example from a global asymmetry due to the different composition of the aqueous compartments inside and outside the closed membranes. We will however not consider the action of proteins that might localize only to specific parts of the membrane, leading to varying values of $m$ throughout the membrane.
8. High surface densities of proteins can influence the elastic moduli $\kappa$ and $\kappa_G$ of the membrane. Atg8-PE, for example, has been shown to induce membrane tubulation and rigidification[15]. This can however be attributed to un-physiological high surface densities of Atg8-PE[14]. As protein coats were not found on autophagosomes thus far [4] the assumption of constant elastic moduli seems reasonable.

## 3 Formation of autophagosomes under consideration of geometric constraints

### 3.1 Geometry and energetics of common membrane shapes

9. The geometry of a spherical **vesicle** is defined by its radius $R_{\text{sp}}$, see Fig. 1. Its surface area is given by $A = 4\pi R_{\text{sp}}^2$, and its enclosed volume by $V = 4\pi R_{\text{sp}}^3/3$. Its mean curvature is given by $M = 1/R_{\text{sp}}$ everywhere, with which we can calculate its bending energy to be $E_{\text{be}} = 8\pi\kappa\left(1 - mR_{\text{sp}}\right)^2 + 4\pi\kappa_G$. Interestingly, in the absence of spontaneous curvature (with $m = 0$), the bending energy of a sphere becomes independent of its radius. Throughout this chapter, vesicles are considered spherical.

10. A cylindrical **membrane tubule** is defined by its radius $R_{cy}$, as well as by its length $L$, see Fig. 1. Assuming that the tubule is capped at both ends, its surface area will be given by $A = 2\pi R_{cy}L + 4\pi R_{cy}^2$, and its volume by $V = \pi R_{cy}^2 L + 4\pi R_{cy}^3/3$. In both cases, the second term corresponding to the area and volume of the two caps becomes negligible if the tubule is very long, i.e. if $L \gg R_{cy}$. The mean curvature along the cylindrical part is $M = 1/2R_{cy}$, and it is $M = 1/R_{cy}$ at the two end caps. With these, we can calculate the bending energy of a tubule to be $E_{be} = \pi\kappa \frac{L}{R_{cy}}(1 - 2mR_{cy})^2 + 8\pi\kappa(1 - mR_{cy})^2 + 4\pi\kappa_G$.

11. A flat **phagophore**, a pancake-like, double-**membrane sheet** can be defined by its long radius $R_{sh}$ and by its rim radius $r_{rim}$, see Fig. 1. Such a definition is equivalent to constructing the pancake as two flat single-membrane disks connected to each other by a toroidal rim. Because the equations for the area, volume and mean curvature of a toroidal segment are rather complicated, it is useful to consider only the case of a very thin sheet, i.e. of $r_{rim} \ll R_{sh}$. In this case, the toroidal segment can be approximated as being half a cylinder of length $L = 2\pi R_{sh}$, and the relevant equations become simpler. The area of the sheet then becomes $A \approx 2\pi R_{sh}^2 + 2\pi^2 R_{sh} r_{rim}$, and the enclosed volume $V \approx 2\pi R_{sh}^2 r_{rim} + \pi^2 R_{sh} r_{rim}^2$. The flat disks of the pancake have $M = 0$ while the rim has $M \approx 1/2r_{rim}$, leading to a bending energy $E_{be} \approx 4\pi\kappa m^2 R_{sh}^2 + \pi^2\kappa \frac{R_{sh}}{r_{rim}}(1 - 2mr_{rim})^2 + 4\pi\kappa_G$. Furthermore, we note that, for sheets that are large but thin, the calculation of their area and volume is particularly simple, even in cases when their shape is not pancake-like. In this case, a sheet (or a collection of sheets) can be defined by its projected area $A_{sh}$ (corresponding to $\pi R_{sh}^2$ for a pancake-like sheet) and its thickness $h$ (corresponding to $2r_{rim}$ in the case of a pancake-like sheet). The total area of the sheet is then $A \approx 2A_{sh}$ and its volume is $V \approx A_{sh}h$.

12. Finally, the shape of an **autophagosome** consists of an inner autophagic vesicle, a sphere of radius $R_{in}$ which is nested inside a slightly larger outer autophagic vesicle, a sphere of outer radius $R_{out}$, see Fig. 1. Alternatively, we could also define the shape of an autophagosome by its outer radius $R_{out}$ and the spacing between the inner and outer spheres $d \equiv R_{out} - R_{in}$. The area of the autophagosome will be $A = 4\pi(R_{out}^2 + R_{in}^2)$, and its lumen or enclosed volume $V = 4\pi(R_{out}^3 - R_{in}^3)/3$. We note that, by enclosed volume, we refer to the volume within the bounds of the continuous membrane that forms the autophagosome prior to scission of the autophagosomal neck, and not the volume captured within the inner vesicle. The mean curvature of the inner segment of the autophagosome is $M = -1/R_{in}$ and is negative because it bulges towards the interior, whereas the mean curvature of the outer segment is $M = 1/R_{out}$. With these, the bending energy of an autophagosome can be calculated as $E_{be} = 8\pi\kappa(1 - mR_{out})^2 + 8\pi\kappa(1 + mR_{in})^2 + 4\pi\kappa_G$. Notice that, in the absence of spontaneous curvature (with $m = 0$), the bending energy of the autophagosome is simply $E_{be} = 16\pi\kappa + 4\pi\kappa_G$, which is independent of the size of the autophagosome or the relative size of its inner and outer radii.

13. **Typical dimensions in cells of the membrane shapes considered here**: 1) $R_{sp}$ of vesicles is in the range of 20-50 nm for Atg9 and COPII vesicles, respectively[16,17]. 2) $R_{cy}$ of ER tubules is about 15 nm in yeast and 30 nm in mammals [9]. 3) The thickness $h$ of ER sheets is about 50 nm[18]. 4) The size of autophagosomes can vary strongly. They can span more than one order of magnitude, from few hundred nanometres to several micrometres[19–21]. The relation between size of autophagosomes and spacing between their inner and outer autophagic vesicles is not well characterized. The spacing is in the order of tens of nanometres[22,23]. Throughout most of this chapter we consider a **typical autophagosome** of $1\,\mu m$ diameter (implying an outer radius

$R_\text{out} = 500$ nm) and intermembrane spacing $d = 20$ nm (implying an inner radius $R_\text{in} = 480$ nm).

## 3.2 Formation of autophagosomes from various pre-autophagosomal membranes

14. **Vesicles**

    Suppose that we observe an autophagosome with outer radius $R_\text{out}$ and inner radius $R_\text{in}$. How many spherical vesicles, of what size, would need to fuse in order to generate such an autophagosome? Because successful membrane fusion will conserve both area as well as enclosed volume, we need to find the number and size of vesicles that will have the same area and volume as the final autophagosome. The conservation of area gives us the condition $A = 4\pi(R_\text{out}^2 + R_\text{in}^2) = N4\pi R_\text{sp}^2$, where $N$ is the number of vesicles, whereas the conservation of volume gives the condition $V = 4\pi(R_\text{out}^3 - R_\text{in}^3)/3 = N4\pi R_\text{sp}^3/3$. Combining both conditions and solving for $N$ and $R_\text{sp}$, we obtain

    $$N = \frac{(R_\text{out}^2 + R_\text{in}^2)^3}{(R_\text{out}^3 - R_\text{in}^3)^2} \quad (2)$$

    $$R_\text{sp} = \frac{R_\text{out}^3 - R_\text{in}^3}{R_\text{out}^2 + R_\text{in}^2} \quad (3)$$

    This highlights a perhaps unintuitive fact: there is only one possible vesicle size that, when combined in a specific number, will lead to an autophagosome of given size and shape (i.e. given spacing between outer and inner membranes). One might have thought that it would be possible to create the same autophagosome from a smaller number of larger vesicles, or from a larger number of smaller vesicles, but this is not the case, due to the fact that both the area as well as the enclosed volume of the autophagosome must be taken into account. One could, of course, use combinations of vesicles of different sizes, but as long as the difference in size between the vesicles involved is small, the typical vesicle size should be given by eq (3). For our example autophagosome of 1 $\mu$m diameter and $d = 20$ nm, eqs (2) and (3) predict that $N = 534$ vesicles of radius $R_\text{sp} = 30$ nm must have fused, see Fig. 2. In Fig. 3, we plot the number and radius of the vesicles required for the generation of an autophagosome with three different outer radii and for varying intermembrane spacing $d$ of the autophagosome.

    The size of the vesicles required is well within the biologically reasonable range, see Note 13. In contrast, hundreds of vesicles were not reported in the context of autophagosome formation so far. This is especially important as autophagosomes are characterized by a small intermembrane spacing. To obtain autophagosomes from vesicles exclusively in this regime, hundreds of vesicles are required to fuse with each other, see Fig. 3a. This is valid for a large range of autophagosomal sizes. Thus, the absence of excessive numbers of vesicles suggests that vesicles might contribute a minor fraction of the final autophagosomal membrane.

    Eqs (2) and (3) become particularly simple when we consider the biologically relevant case of very small intermembrane spacing of the autophagosome, i.e. $d \ll R_\text{out}$. In this case, a series expansion of (2) and (3) in powers of $d$ gives $R_\text{sp} \approx (3/2)d$ and $N \approx (8/9)(R_\text{out}/d)^2$. In this limit, therefore, the diameter of the fusing vesicles must be three times the intermembrane spacing of the final autophagosome, and the number of vesicles necessary grows quadratically with the ratio of outer radius to intermembrane spacing of the autophagosome. Importantly, this limit of $d \ll R_\text{out}$ is the typical case *in vivo*.

15. **Membrane tubules**

It is also possible that the membrane used to generate autophagosomes has its origin in membrane tubules such as those in the ER. What is then the total length and diameter of tubules needed to generate a particular autophagosome? Assuming that the tubules involved are long and thin (which, as we will see *a posteriori*, corresponds to autophagosomes with narrow intermembrane spacing), we can neglect the area and volume of the tubule end caps, and the conservation of area and volume during formation of the autophagosome gives us the conditions $A = 4\pi(R_{\text{out}}^2 + R_{\text{in}}^2) = 2\pi R_{\text{cy}} L$ and $V = 4\pi(R_{\text{out}}^3 - R_{\text{in}}^3)/3 = \pi R_{\text{cy}}^2 L$. We note that these two conditions are valid independently of whether we consider a single tubule of length $L$, or several separate shorter tubules that fuse, as long as their total length adds up to $L$. Taking both conditions together and solving for $L$ and $R_{\text{cy}}$, we find

$$L = 3\frac{(R_{\text{out}}^2 + R_{\text{in}}^2)^2}{R_{\text{out}}^3 - R_{\text{in}}^3} \quad (4)$$

$$R_{\text{cy}} = \frac{2}{3}\frac{R_{\text{out}}^3 - R_{\text{in}}^3}{R_{\text{out}}^2 + R_{\text{in}}^2} \quad (5)$$

As in the case of spherical vesicles, there is only one specific combination of total length and radius of tubules that can give rise to an autophagosome of given geometry. In order to generate our typical autophagosome of 1 $\mu$m diameter and $d = 20$ nm, eqs (4) and (5) predict that a total length $L = 48\ \mu$m of tubules of radius $R_{\text{cy}} = 20$ nm is required, see Fig. 2.

Interestingly, comparing eqs (3) and (5) we see that the required radius of the cylindrical tubules involved is two thirds smaller than the radius required for the vesicles. In particular, if we once again consider the case of very small intermembrane spacing of the autophagosome ($d \ll R_{\text{out}}$), eqs (4) and (5) simplify to $L \approx 4R_{\text{out}}^2/d$ and $R_{\text{cy}} \approx d$. Therefore, the diameter of the cylindrical tubules is generally twice the intermembrane spacing, and the total length of tubules required grows quadratically with the size of the autophagosome (if the intermembrane spacing is kept constant).

16. **Membrane sheets**

The ER is also composed of membrane sheets, and it is possible that these participate in the generation of autophagosomes. As before, the conservation of membrane area and enclosed volume during autophagosome formation can be used to calculate the total amount and the thickness of the sheets involved in the process. Assuming that the sheets involved are large but thin, area and volume conservation gives the conditions $A = 4\pi(R_{\text{out}}^2 + R_{\text{in}}^2) = 2A_{\text{sh}}$ and $V = 4\pi(R_{\text{out}}^3 - R_{\text{in}}^3)/3 = A_{\text{sh}} h$. As in the case of the tubules, these conditions are valid independently of whether we consider a large sheet of projected area $A_{\text{sh}}$, or several smaller sheets that fuse, as long as their total projected area adds up to $A_{\text{sh}}$. Taking both conditions together and solving for $A_{\text{sh}}$ and $h$, we obtain

$$A_{\text{sh}} = 2\pi(R_{\text{out}}^2 + R_{\text{in}}^2) \quad (6)$$

$$h = \frac{2}{3}\frac{R_{\text{out}}^3 - R_{\text{in}}^3}{R_{\text{out}}^2 + R_{\text{in}}^2} \quad (7)$$

For our typical autophagosome of 1 $\mu$m diameter and $d = 20$ nm, eqs (6) and (7) predict that a total projected area $A_{\text{sh}} = 3.0\ \mu\text{m}^2$ of sheets of thickness $h = 20$ nm is required. Such a total projected area would be equivalent to a single pancake-like sheet with diameter $2\sqrt{A_{\text{sh}}/\pi} = 2.0\ \mu$m, see Fig. 2.

As in the case of spheres and tubules, eqs (6) and (7) become particularly simple for the case of an autophagosome with very small intermembrane spacing ($d \ll R_{\text{out}}$). In this case, they become $A_{\text{sh}} \approx 4\pi R_{\text{out}}^2$ and $h \approx d$. The required thickness of the sheets is therefore approximately equal to the intermembrane spacing of the final autophagosome, and the total projected area of the sheets required grows quadratically with the size of the autophagosome.

17. **Mixtures of vesicles, tubules, and sheets**
   In short, we have found above that an autophagosome with intermembrane spacing $d$ may be formed from many vesicles of diameter $3d$, from one or multiple long tubules of diameter $2d$, or from one or multiple membrane sheets of thickness $d$, see Fig. 2. Moreover, the autophagosome could be formed from any mixture of such vesicles, tubules, and sheets: as long as their total area equals the total area of the autophagosome, the enclosed volume will automatically have the correct value.
   When the autophagosome is generated from mixtures of these three components, it also becomes possible for the diameter/thickness of the components to be different from the values calculated above. However, if some of the components are smaller than the values calculated above, the rest of the components must be larger. As an example, it is impossible to generate an autophagosome of intermembrane spacing $d$ from a mixture of vesicles of diameter smaller than $3d$ and tubules of diameter smaller than $2d$. If the vesicle diameter is smaller than $3d$, the tubule diameter will need to be larger than $2d$, and *vice versa*.

18. **In cells**, it seems rather unreasonable to assume that the membrane source of autophagosomes is controlled precisely to account for the area-volume constraint, see Note 17. In the case that one of the membrane sources were too large, excess luminal volume (relative to the membrane area) may accumulate during the formation of autophagosomes. Because autophagosomes are known for a small luminal volume[22], this raises the question of how can the cell correct for such events.
   The membrane of a developing autophagosome can be continuous with other organelles such as the sheets of the ER[24–26]. This implies that the lumen of the sheets and lumen of the autophagosome may be continuous as well and might be used to transfer luminal volumes between ER and autophagosomes. Interestingly, the luminal spacing of ER sheets is significantly larger than that of the autophagosome connected to it[25,26]. This suggests that autophagosomes are either more likely to be grown from tubular ER than sheet-like ER, and/or that a mechanism exists which supplies membranes with a disproportionately large area-to-volume ratio.

# 4 Energetics of autophagosome formation

We have shown so far that, because membrane area and enclosed volume are conserved during autophagosome formation, there are important constraints on the number, shape, and dimensions of the precursor membranes (vesicles, sheets, tubules) participating in the process. A second important question concerns the energetics of the process: under which conditions is autophagosome formation favourable energetically? Is the bending energy of the system reduced or increased during autophagosome formation?

## 4.1 Autophagosome-like shapes can form by fusing three vesicles

19. Straightforward considerations can be made to show that autophagosome formation is generally an energetically favourable process. In ref. 12 the energy of closed vesicles of different shapes was studied in detail, as a function of their volume-to-area ratio $v \equiv V/[(4\pi/3)(A/4\pi)^{3/2}]$, which can range between 0 and 1, with $v = 1$ corresponding to a sphere. For the case of membranes with zero spontaneous curvature, it was found that tubule-like (prolate) shapes have lowest energy in the range $1 > v > 0.652$, sheet-like (oblate) shapes have lowest energy in the range

$0.652 > v > 0.592$, and autophagosome-like (stomatocyte) shapes have lowest energy in the range $0.592 > v > 0$. A straightforward but important consequence of this is the following. When two, three, or four spherical vesicles fuse, the volume-to-area ratio of the resulting post-fusion vesicle is 0.707, 0.577, and 0.5, respectively. The fusion of even larger numbers of vesicles result in even lower volume-to-area ratio of the final vesicle. Taking these values together with the ranges for lowest energy shapes described above, one concludes that fusion of three or more vesicles is sufficient for autophagosome-like shapes to be energetically preferred over sheet-like or tubule-like shapes. This was recently confirmed by computer simulations[27]. Note: The volume-to-area ratio of our typical autophagosome of 1 $\mu$m diameter and $d = 20$ nm is $v = 0.043$.

In order to fully make sure that the lowest possible energy corresponds to the autophagosome-like shape, we should also compare this energy to the total energy of the individual vesicles before fusion. As described in Section 3.1, in the absence of spontaneous curvature, the energy of $N$ spherical vesicles is $E_{\text{ve}} = N[8\pi\kappa + 4\pi\kappa_G]$. The energy of the resulting autophagosome after fusion, on the other hand, is $E_{\text{ap}} = 16\pi\kappa + 4\pi\kappa_G$. The condition that the autophagosome shape has lower or equal energy, i.e. that $E_{\text{ap}} \leq E_{\text{ve}}$, is then equivalent to

$$N \geq \frac{4\kappa + \kappa_G}{2\kappa + \kappa_G} \quad (8)$$

For a typical membrane, we have $\kappa_G \approx -\kappa$, which results in $N \geq 3$. Therefore, fusion of three or more vesicles into an autophagosome-like shape can decrease the energy of the system indeed. If the Gaussian curvature modulus $\kappa_G$ were smaller or larger than $-\kappa$, this minimal number of vesicles would increase or decrease, respectively.

### 4.2 Relaxation of total bending energy during autophagosome formation

The calculations described above show that, already for autophagosomes formed from as little as three vesicles, the formation process can be energetically favourable. One may wonder then, how large is the relaxation of bending energy during the formation of a realistic autophagosome for which, as we saw above, hundreds of vesicles (or a comparable amount of membrane in the form of tubules or sheets) must fuse?

20. In Section 3, we obtained the number and radius of vesicles, length and radius of tubules, or size and thickness of membrane sheets, that are needed to generate an autophagosome of given size and intermembrane spacing. It is straightforward to combine these results with the expressions for the bending energy of each of these shapes that we derived in Section 2.3, and in this way be able to compare the total bending energy of each of these four configurations.
In Figure 4a, we plot the bending energy of each of these configurations for the particular case of zero spontaneous curvature of the membrane, for an autophagosome with outer radius $R_{\text{out}} = 500$ nm, as a function of the intermembrane spacing $d$ of the autophagosome. We find that, for biologically reasonable values of the intermembrane spacing, the ordering of the different configurations from highest to lowest energy is long tubule, set of vesicles, membrane sheet, and autophagosome. The total relaxation of bending energy when a tubule becomes an autophagosome can be extremely large, of the order of $10^3 \kappa \approx 10^4 k_B T \approx 10^{-16}$ J. This energy is larger for smaller intermembrane spacings of the autophagosome.

### 4.3 Pathways of autophagosome formation
Autophagosomes form by a complex cascade of membrane remodeling events[1]. In general, such remodelling processes can be described as a combination of changes in membrane morphology and topology[28–30]. Morphological transformations arise from continuous and smooth changes of the membrane curvature and shape, whereas topological transformations of the membrane involve intermediate states in which the membranes have to deviate strongly from their usual bilayer structure.

21. **Topological transformations**, membrane fusion or scission, are required to change the shape of membrane between vesicles and other shapes such as tubules, see Fig. 1. The total energy of the tubules and vesicles is in a similar range, see Fig. 4a. Alterations in membrane topology include rather complicated energy barriers with non-bilayer transition states. Despite the fact that vesicles are known to be involved in autophagosome formation[6], their total number required *in vivo* seems to be much lower than estimated ($N > 500$), see Note 14. For these reasons, this transition will not be considered in what follows.

22. **Morphological transformations** could continuously transform a long tubule into a large sheet or an autophagosome without requiring membrane fusion/fission. This transition can in principle occur in a single discrete step, involving full-scale shape transformation of the tubule into a sheet *via* intermediate paddle-like shapes of lower symmetry, which represent a sizable energy barrier[27,31,32]. Alternatively, one could consider a process that begins after the nucleation (or fusion) of a small sheet or autophagosomal structure at one end of a long tubule. This small sheet or autophagosomal structure can then continuously grow into a large sheet or mature autophagosome by exchanging area and volume with the tubule, which must then continuously shrink as a consequence. In order to explore such pathways, we consider the energy of the system when the total area and volume (which correspond to a 'mature' autophagosome of given size and shape) is partitioned into a single, continuous membrane. Various membrane fractions represent a sheet of changing size, a tubule of varying length, and a 'growing' autophagosome. The radius of the vesicles and the tubule are given by eqs (3) and (5). In Figure 4b, we plot the energy of such a mixture for the particular case of a target autophagosome with $R_{\text{out}} = 500$ nm and $d = 20$ nm. The corners of the ternary diagram correspond to the target autophagosome (top corner), the case with only a sheet (bottom left corner), and the long tubule only (bottom right corner). The lines directly connecting the corners of the diagram represent mixtures between two of the shapes. As we had seen before, the tubule has highest energy, whereas the 'mature' autophagosome has lowest energy. All pathways going through the inside of the ternary diagram are more complex and involve the simultaneous coexistence of three membrane shapes: sheet, tubule, and autophagosome.

23. Finally, let us briefly consider the **effect of spontaneous curvature** on the energetics of autophagosome formation. For zero spontaneous curvature, we found the ordering of tubule, vesicles, sheet, and autophagosome from larger to lower energy. Would this ordering change for non-zero spontaneous curvature? In Fig. 5, we plot the energies of the four configurations as a function of spontaneous curvature, for the particular case of an autophagosome with outer radius $R_{\text{out}} = 500$ nm and intermembrane spacing $d = 20$ nm. Indeed, we find that this ordering can change, and in particular vesicles and tubules become more energetically favourable for sufficiently high spontaneous curvature. This was to be expected, given that for example proteins inducing positive spontaneous curvature stabilize tubular structures in the ER[33]. Nevertheless, we find that autophagosomes are always energetically more stable that membrane sheets of the corresponding size, which is in agreement with previous work considering the bending of phagophores[4]. Moreover we show that the ordering of tubule, vesicles, sheet, and autophagosome from larger to lower energy survives for a wide range of positive and negative spontaneous curvatures around $m = 0$.

**Acknowledgement:** We thank Reinhard Lipowsky (MPI of Colloids and Interfaces) for stimulating discussions, institutional and financial support.

**References (Arabic numbers should be used for text citations, will be changed later)**

**Figures**

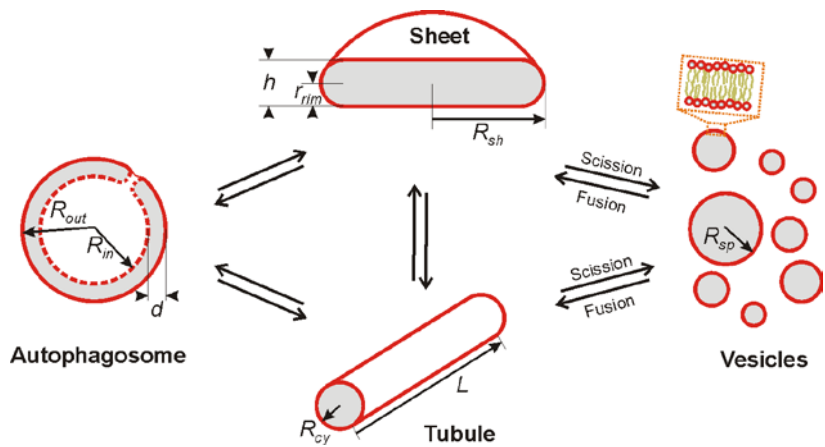

**Figure 1. Mayor shapes of membranes during autophagy and their geometric parameters.** In principle, autophagosomes can be generated from several membranous precursors such as sheets, tubules, fused vesicles and mixtures of various sources. 2-dimensional views show membranes in red as indicated. Dashed lines and grey areas refer to membranes and enclosed volumes which are visible only in sections. Included are the parameters which define each of the shapes and are used in this chapter.

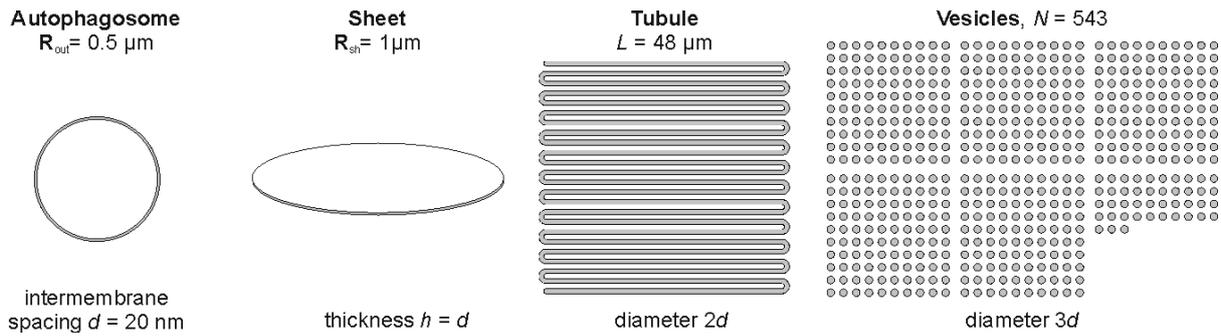

**Figure 2: Geometries of autophagosomal membranes.** Precursor membranes are drawn to scale and match the membrane area and internal volume of our standard autophagosome with $R_\text{out} = 500$ nm and intermembrane spacing $d = 20$ nm. Characteristic length scales and vesicle numbers are given. Diameters of the sheet's rim, the tubule and the vesicles are given relative to the intermembrane spacing of the autophagosome $d$.

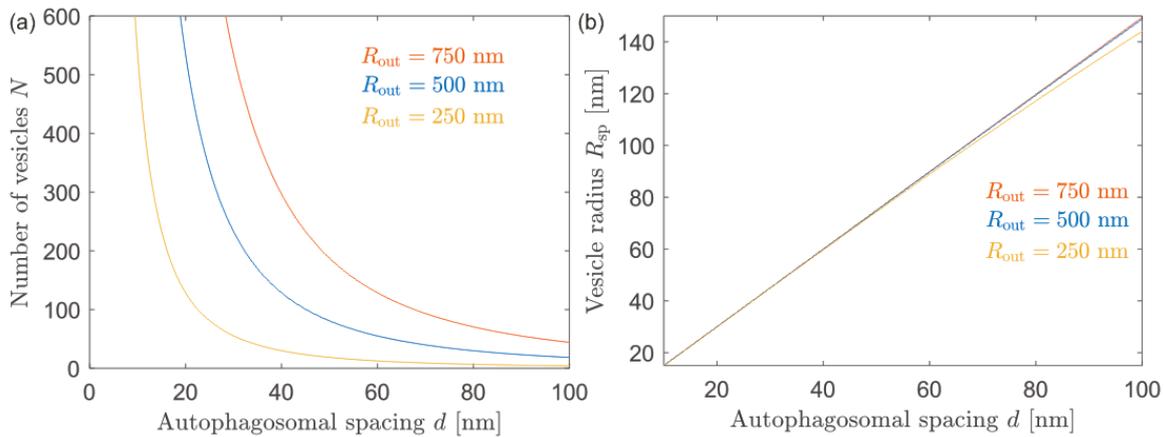

**Figure 3: Generation of an autophagosome from fusion of spherical vesicles.** (a) Number of vesicles and (b) radius of the vesicles required for the generation of autophagosomes with outer radii $R_\text{out} = 750$ nm, 500 nm, and 250 nm (i.e. diameter 1.5 $\mu$m, 1 $\mu$m, and 500 nm), as a function of the intermembrane spacing $d$ of the autophagosome. These are calculated from eqs (2) and (3) in the text. The smaller the spacing $d$, a larger number of smaller vesicles is required.

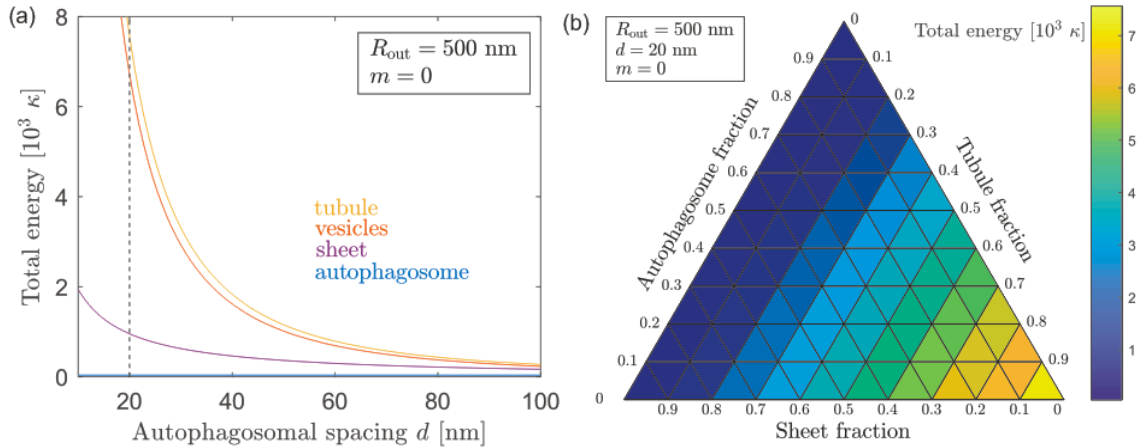

**Figure 4: Bending energy relaxation during autophagosome formation in the absence of spontaneous curvature.** (a) Bending energy of an autophagosome with outer radius $R_\text{out} = 500$ nm (blue), as well as the corresponding single sheet (purple), set of $N$ vesicles (red, Fig. 3a), and single tubule (yellow) that have the same total area and enclosed volume as the autophagosome. Within a realistic range of autophagosomal spacing $(10 - 100 \text{ nm})$, we find that the autophagosome has lowest energy, followed by the sheet, the vesicles, and the tubule, in this order. The vertical dashed line corresponds to the particular case discussed throughout the chapter, with intermembrane spacing $d = 20$ nm. (b) Bending energy of a sheet, a long tubule, and an autophagosome, such that the total membrane area and enclosed volume of the mixture corresponds to an autophagosome with outer radius $R_\text{out} = 500$ nm and intermembrane spacing $d = 20$ nm. The top, bottom left, and bottom right corners of the triangle therefore correspond to the special cases of only autophagosome, only sheet, and only tubule. The energy values at these corners are thus identical to the values of the blue, purple, and yellow lines in (a) at $d = 20$ nm.

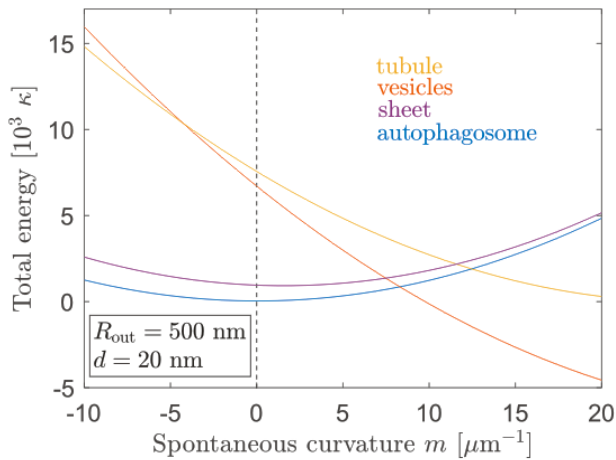

**Figure 5: Bending energy relaxation during autophagosome formation as a function of spontaneous curvature.** The bending energy of an autophagosome with outer radius $R_\text{out} = 500$ nm and intermembrane spacing $d = 20$ nm (blue), as well as the bending energies of the corresponding single sheet (purple), set of $N = 534$ vesicles (red), and single tubule (yellow) that have the same total area and enclosed volume as the autophagosome, are plotted as a function of spontaneous curvature. Spontaneous curvature can strongly affect the energy of the different configurations, but the ordering of tubule, vesicles, sheet and autophagosome from highest to lowest bending energy remains valid in a wide range of spontaneous curvatures above and below $m = 0$ (vertical dashed line). Formation of autophagosomes from sheet-like membranes, such as phagophores, is always energetically favourable.